\documentstyle[11pt,newpasp,twoside,epsf]{article}
\def\edcomment#1{\iffalse\marginpar{\raggedright\sl#1\/}\else\relax\fi}
\reversemarginpar

\begin{document}
\title{Contact Binaries}
\author{Ronald F. Webbink}
\affil{Department of Astronomy, University of Illinois, 1002 W. Green St.,
Urbana, IL 61801, U.S.A.}

\begin{abstract}
Despite being the most abundant (by space density) of interacting binary
stars, W~Ursae Majoris (W~UMa) stars are not understood structurally.  Their
stellar components are in physical contact, and share a common convective
envelope.  Observations demand large-scale energy exchange between components,
as do considerations of hydrostatic and thermal equilibrium.  Yet solutions in
complete equilibrium can account for only a few of the bluest W~UMa binaries. 
Models which permit departures from thermal equilibrium can reproduce the
observable properties of later-type W~UMa binaries, but they develop
relaxation oscillations, during a substantial portion of which the binary
components break thermal contact and develop very different effective
temperatures, contrary to observational statistics.  Massive, early-type
contact binaries are also known to exist, but the nature of energy and mass
exchange in common radiative envelopes remains virtually unexplored.
\end{abstract}

\section{Introduction}

In close binary systems, the most extreme examples of tidal distortion are
found observationally among contact binary stars, the most common variety
being the W~Ursae Majoris (W~UMa) variables.  The apparent brightnesses of
these systems are continuously variable, even out of eclipse, a property which
signals the breakdown of symmetry of the component stars about their
respective rotational axes.  As illustrated in Figure~1 by the eponymous
variable W~UMa itself, W~UMa variables are characterized as well by having
primary and secondary eclipses of nearly equal depths.  In addition, they show
very little color variation through eclipse.  Since the ratio of eclipse
depths is dictated by the ratio of surface brightnesses of the component
stars, the near equality of eclipse depths and absence of pronounced color
variation in either eclipse demand that the component stars are nearly
identical in effective temperature, $T_1 \approx T_2$.  Detailed light curve
synthesis models, employing Roche potentials for the equilibrium surfaces of
the two stars, show that the component stars in W~UMa binaries exist in a
state of physical contact.

\begin{figure}[t]
\plotone{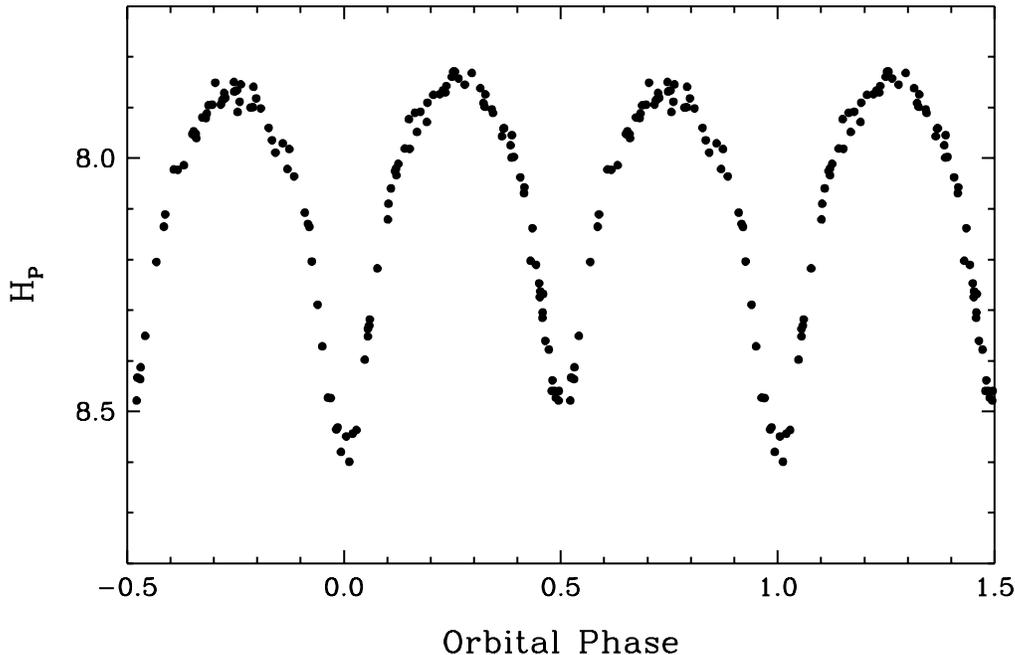}
\caption{The light curve of W Ursae Majoris observed by \emph{Hipparcos}.}
\end{figure}
In orbital period, the W~UMa binaries are overwhelmingly concentrated at short
orbital period, down to $P = 5\fh3$ (CC Com), with few systems at periods $P >
16^{\rm h}$.  Because both stars fill their Roche lobes, the orbital periods
of W~UMa systems are closely coupled to the stellar mean densities, with $P
\approx 8^{\rm h}\,(\overline{\rho}/\overline{\rho}_{\odot})^{1/2}$.  The
orbital period range occupied by W~UMa systems then immediately tells us that
they reside on or near the main sequence, with primary masses roughly in the
range 0.6 to 2 M$_{\odot}$.  That inference is reinforced by a strong
correlation between orbital period and dereddened color (Eggen 1967).  These
are therefore systems in which the primary components are still in core
hydrogen burning.  The fact that W~UMa systems are found in clusters along
their main sequences, often below the cluster turn-off (\emph{e.g.}, Rucinski
1998), confirms this conclusion, and implies that the secondary components as
well are in core hydrogen burning.

Spectroscopic studies of W~UMa systems reveal that, despite the facts that
these binaries reside along the main sequence and that their component
effective temperatures a nearly equal, they invariably have unequal masses, in
some cases spectacularly so.  The Roche geometry requires that the ratio of
radii of contact components vary as $R_2/R_1 \approx (M_2/M_1)^{0.46}$ (Kuiper
1941), whereas the ratio of radii of single low-mass zero-age main sequence
stars depends more strongly on mass, $R_2/R_1 \approx M_2/M_1$.  The
near-equality of effective temperatures in W~UMa binaries therefore demands
that the ratio of radiated luminosities vary as $L_2/L_1 \approx
(M_2/M_1)^{0.92}$, whereas the ratio of nuclear luminosities for two zero-age
main sequence stars of the same respective masses varies roughly as
$(L_2/L_1)_{\rm nuc} \approx (M_2/M_1)^{4.5}$ at the low masses appropriate to
these binaries.  Among well-studied W~UMa systems, one finds within
observational uncertainties that $(L_1 + L_2) \approx (L_{\rm 1,nuc} + L_{\rm
2,nuc})$ (\emph{e.g.}, Mochnacki 1981), so it is evident that the secondary
(less massive) component is drawing much of its radiant luminosity from the
core of the primary (more massive) component.  This energy transfer amounts to
a relatively modest reduction in the luminosity of the primary, but clearly
initiates a major restructuring of the secondary, which becomes oversized and
overluminous for its mass.  Delineating the mechanism which effects energy
transfer, and accounting for its stability (or lack thereof), are the core
problems in understanding the structure and evolution of W~UMa systems.

W~UMa binaries come in two flavors, A-type and W-type.  Observationally, they
are distinguished by whether the primary (deeper) eclipse corresponds to
transit of the (less massive) secondary across the face of the primary
(A-type), or to occultation of the secondary by the primary (W-type).  There
is a fairly clean separation of the subtypes by spectral type: A-type W~UMa
systems have A-F spectra, and W-type systems have G-K spectra.  The difference
between primary and secondary eclipse depths is quite small (amounting to no
more than a few hundredths of a magnitude), but reflects a small difference in
the eclipsed surface brightnesses of the two components.  In the A-type
systems, this difference can be laid to the effects of gravity-darkening in an
otherwise homogeneous common envelope (the transiting secondary having a
slightly lower mean gravity, and hence lower surface brightness).  In the
W-type systems, a small but real difference in mean surface brightnesses
between components is indicated, with the secondary's surface actually
slightly brighter than the primary's.  Formally, this difference reflects a
difference in effective temperatures (in the sense that effective temperature
is defined in terms of bolometric surface brightness), but that interpretation
is somewhat misleading, in that eclipse depths in the W-type systems remain
nearly constant with wavelength through the near ultraviolet (\emph{e.g.},
Eaton, Wu and Rucinski 1980), implying that the \emph{color} temperatures of
the two stars are nearly identical.  Preferential starspot formation on the
primary components of W-type systems (Figure~2) may provide an explanation for
the lack of wavelength dependence in relative eclipse depths (Rucinski 1985,
1993), if not the underlying difference in effective temperatures.

\begin{figure}[t]
\plotone{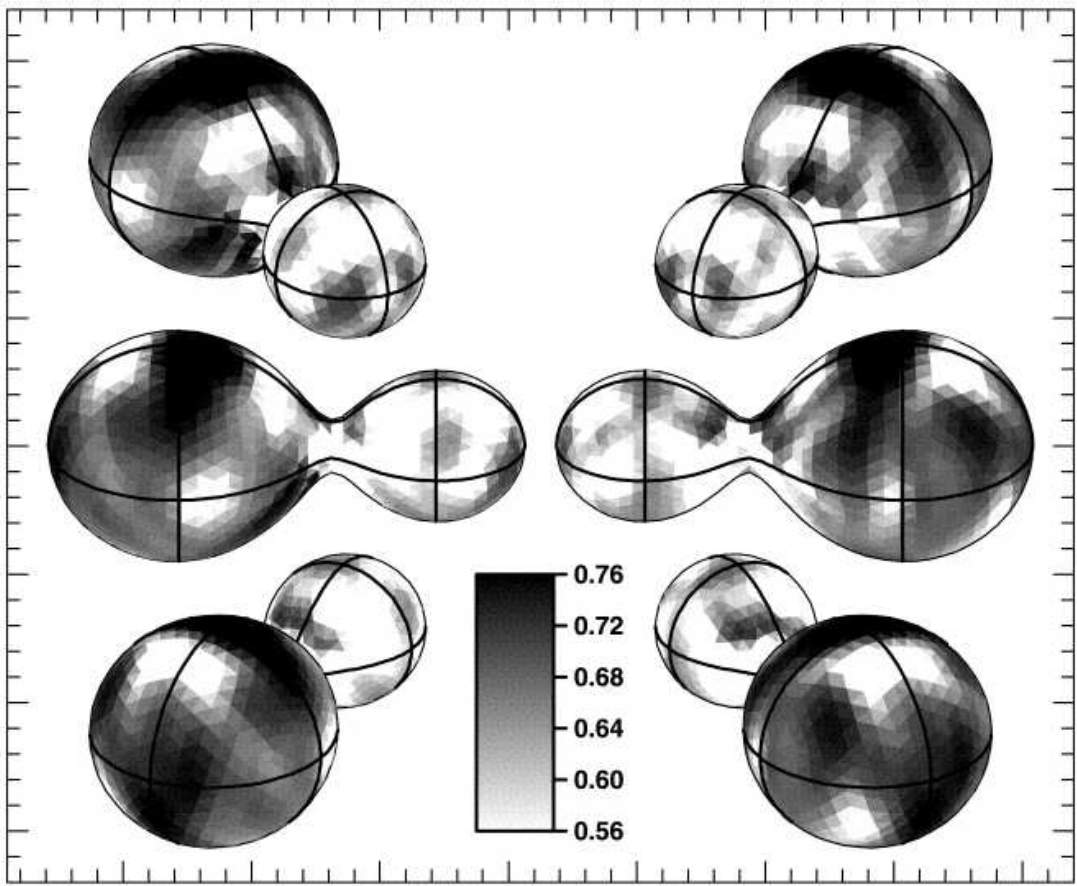}
\caption{Doppler image of VW Cephei for 1992 November/December (Hendry \&
Mochnacki 1981).  The gray scale signifies fractional spot coverage. 
(\copyright American Astronomical Society, by permission)}
\end{figure}

Since the first serious attempt to estimate the space density of W~UMa
binaries (Shapley 1948), it has been evident that these are extremely common
objects.  In gravitational microlensing surveys of the galactic bulge,
foreground W~UMa variables are easily the most common periodic variables
detected.  (They are too faint, intrinsically, to be picked up in great
numbers in the Magellanic Clouds.)  Rucinski (1998) estimated from the OGLE~I
results that they account for roughly 1/130 of solar-type stars in the
foreground disk projected against the galactic bulge.\footnote{In a recent
analysis of the local density of W~UMa systems, based on \emph{Hipparcos}
photometry, Rucinski (2002) finds a significantly lower relative frequency of
occurrence --- roughly 1/500 solar-type stars.  The difference in these
estimates is not yet satisfactorily understood.}  If we take this relative
frequency at face value, it implies that the lifetime of the W~UMa state must
exceed $\sim\!10^8$ yr.  Indeed, the paucity of near-contact \emph{detached}
low-mass binaries to replenish the W~UMa population implies that this is a
very weak lower limit to their lifetimes.  We shall have more to say on the
significance of semi-detached and detached analogues of W~UMa systems below. 
For the present, we note that the high abundance of W~UMa systems requires
that they survive for many times the thermal time scales of their primary
components.

\section{Thermal Equilibrium Models}

In a uniformly-rotating binary star, the acceleration of a test particle can
be derived from an effective potential $\Psi$ (the Roche potential in the
limit that each binary component can be approximated by a point mass at its
center).  The condition of complete hydrostatic equilibrium then requires that
\begin{equation}
\mathbf{\nabla} \mathit{P} = - \rho \mathbf{\nabla} \mathrm{\Psi} ~ .
\end{equation}
Since the gradient in pressure $P$ coincides in direction with the gradient in
potential $\Psi$, it follows that pressure must be a function of $\Psi$ alone,
$P = P(\Psi)$.  Similarly, the density $\rho$ must also be a function of
$\Psi$ alone:
\begin{equation}
\rho = \frac{{\rm d} P(\Psi)}{{\rm d} \Psi} = \rho(\Psi) ~ .
\end{equation}
Assuming a homogeneous composition on equipotential surfaces, it then follows
that \emph{all} state variables are constant on equipotential surfaces,
\emph{i.e.}, functions of $\Psi$ alone (von Zeipel 1924).  The near-uniformity
of local effective temperature over the surface of a W~UMa binary can then be
understood as a consequence of hydrostatic equilibrium obtaining throughout
its common envelope (Osaki 1965; Lucy 1968).  Energy exchange between contact
components is thus implicitly demanded by hydrostatic equilibrium.

The question remains whether a contact state of hydrostatic equilibrium can be
achieved in a state of thermal equilibrium as well.  The observational fact
that W~UMa binaries inhabit later spectral types implies that they have
convective envelopes.  Efforts to build structural models have therefore
generally assumed that the respective component envelopes can be characterized
by a common specific entropy deep into their envelopes.  These models take
advantage of the fact that stellar convection in low-mass main sequence stars
is so efficient that removing half the luminosity carried by convection (or
increasing it ten-fold) makes scarcely any difference to the run of
temperature with pressure through the envelope, unless transfer occurs in the
superadiabatic part of the convection zone.  In his pioneering study, Lucy
(1968) first applied this assumption to build models of contact binaries in
thermal equilibrium.

\begin{figure}[t!]
\plotone{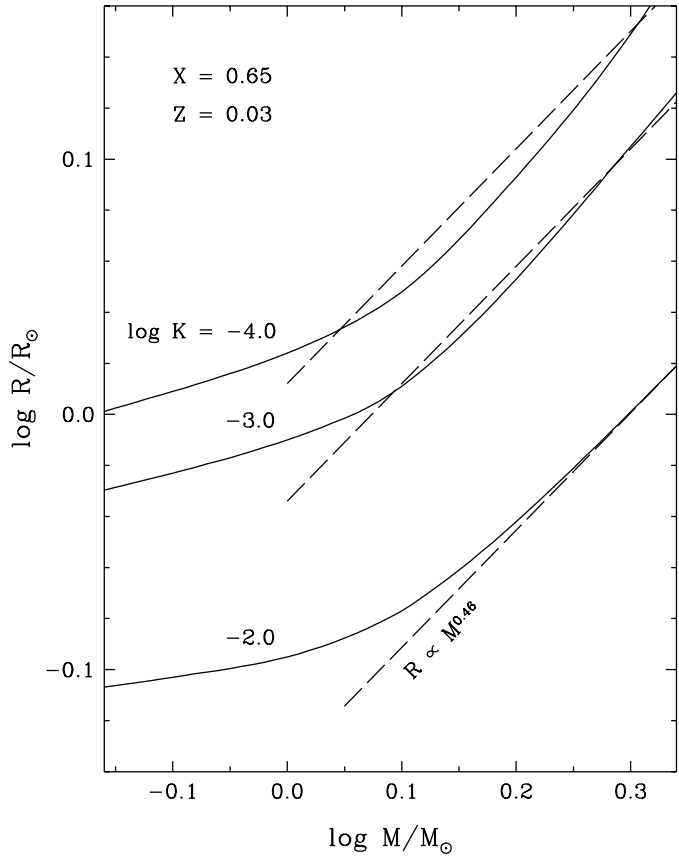}
\caption{The mass-radius diagram for binary components of W~UMa binaries,
after Lucy (1968).  Solid lines mark model sequences with a fixed adiabatic
constant $K$.  Thermal equilibrium solutions are possible where a line
parallel to the Roche lobe mass-radius relation, indicated by a dashed line,
intersects a constant-$K$ curve at two points, and the sum of the luminosities
radiated by the two components equals the sum of their nuclear luminosities. 
The vertical zero point for the Roche lobe mass-radius relation fixes the
binary separation.}
\end{figure}
Lucy's solution technique is illustrated graphically in Figure~3.  He
constructed model stellar sequences in which a common adiabatic constant ($K
\equiv P/T^{5/2}$) characterizing the surface convection zone in both stars,
$K_1 = K_2$, and total blackbody luminosity, $L = 4\pi (R_1^2 + R_2^2) \sigma
T^4$, were substituted as surface boundary conditions for application of the
blackbody law separately to each component.  Thermal equilibrium contact
solutions occur where, for contact components sharing a common value of $K$,
the ratio of stellar radii satisfies the contact condition, and the total
radiated luminosity equals the sum of the respective nuclear luminosities. 
For a given primary mass, solutions of this type are possible only if one
treats the mass of the secondary (or, equivalently, the total orbital angular
momentum) as an eigenvalue of the problem.  Lucy succeeded in finding thermal
equilibrium solutions for zero-age contact binaries with unequal components,
but only in circumstances where the primary component was massive enough to
burn hydrogen in the CN-cycle, while the less massive secondary subsisted on
the $pp$-chain.  That transition between hydrogen-burning networks coincides
with the change in slope of $R(M)$ at fixed $K$ seen in Figure~3.  

In reality, most W~UMa systems are cool and dense enough that both components
must burn hydrogen on the $pp$-chain.  Lucy's thermal equilibrium models were
therefore able to account only for systems of the highest mass, with the
bluest colors, and could not be pushed to lower masses and later spectral
types without appealing to unphysically high CN-cycle reaction rates or
extremely metal-rich compositions (Moss \& Whelan 1970).  Allowance for
evolved primary components permits a modest expansion in the span of solutions
(Hazlehurst 1970), but it cannot account for the numerous relatively red
systems found well below turnoff in old clusters ($cf.$ Rucinski 1998). 
Relaxing the equal-entropy condition for the component envelopes permits one
to build thermal equilibrium models throughout the observed period-color plane
(Biermann \& Thomas 1972, 1973), but these models no longer satisfy the
conditions for hydrostatic equilibrium in the common envelope, and predict
effective temperature differences between components far exceeding those
allowed by the observed small differences between primary and secondary
eclipse depths.

As discussed by many authors (see the review by Smith 1984 for a particularly
lucid account), the problem of building thermal equilibrium models of W~UMa
binaries is over-constrained.  In a detached binary, one can choose the masses
and compositions of the stars independently; interaction terms are small, and
the structure solutions differ little from those of isolated stars.  In a
contact binary, however, hydrostatic equilibrium imposes two additional
constraints: the radii of the two stars must conform to a common
equipotential, and furthermore (appealing to the more tractable convective
envelope case) their envelope entropies must be equal.  For a given total mass
and total angular momentum, the binary mass ratio and the depth of contact
must be eigenvalues of the problem, and solutions are possible only for a very
small range of parameters\footnote{In a controversial series of papers, Shu,
Lubow, \& Anderson (1976, 1979; Lubow \& Shu 1977, 1979) proposed that a
discontinuity in temperature exists at the base of the common envelope, where
it encompasses the less massive component (or more massive component when the
common envelope is radiative).  This contact discontinuity (DSC) hypothesis
makes it formally possible to construct contact binaries in hydrostatic and
thermal equilibrium (except for a singular flux divergence at the contact
discontinuity!) for any combination of masses and any depth of contact
permitted by the Roche potentials, but has been widely criticized as
unphysical (Hazlehurst \& Refsdal 1978; Lucy \& Wilson 1979; Papaloizou \&
Pringle 1979; Smith, Robertson, \& Smith 1980; Eggleton 1981; see also the
reply by Shu, Lubow, \& Anderson 1980).  The introduction of a temperature
jump at the contact discontinuity without an additional constraint removes a
boundary condition where the stellar interior is fit to the common envelope,
leaving the model underconstrained.  As first noted by Lucy \& Wilson, the
evolution of such a model is then indeterminate.}.

\section{Thermal Relaxation Oscillations}

If conditions for thermal equilibrium and hydrostatic equilibrium cannot be
satisfied simultaneously, then given the enormous difference in their
characteristic global time scales (roughly a factor of $10^{11}$ in W~UMa
binaries), it is logical to abandon the requirement for thermal equilibrium. 
Lucy (1976) and Flannery (1976) showed that models of late-type contact
binaries with isentropic common envelopes could then be constructed, but that
such systems suffer cyclic thermal relaxation oscillations (TRO's --- see also
K\"{a}hler, Matraka, \& Weigert 1986$a,b$, 1987).  In this cycle, the system
oscillates between contact and semi-detached states.  In the contact state,
energy transfer from primary to secondary inflates the secondary (driving mass
transfer from secondary to primary).  In the semi-detached state, the primary
continues to fill its Roche lobe, but the secondary initially relaxes from its
energy-transfer-inflated state toward thermal equilibrium, only to be driven
by continuing mass transfer back into a contact state.  As illustrated in
Figure~4, this cycle is driven by the fact that the contact condition demands
of the secondary a radius intermediate between two thermal equilibrium states:
one contact (with an isentropic common envelope) and one detached (with no
mass or energy input from the primary).

\begin{figure}[t!]
\plotone{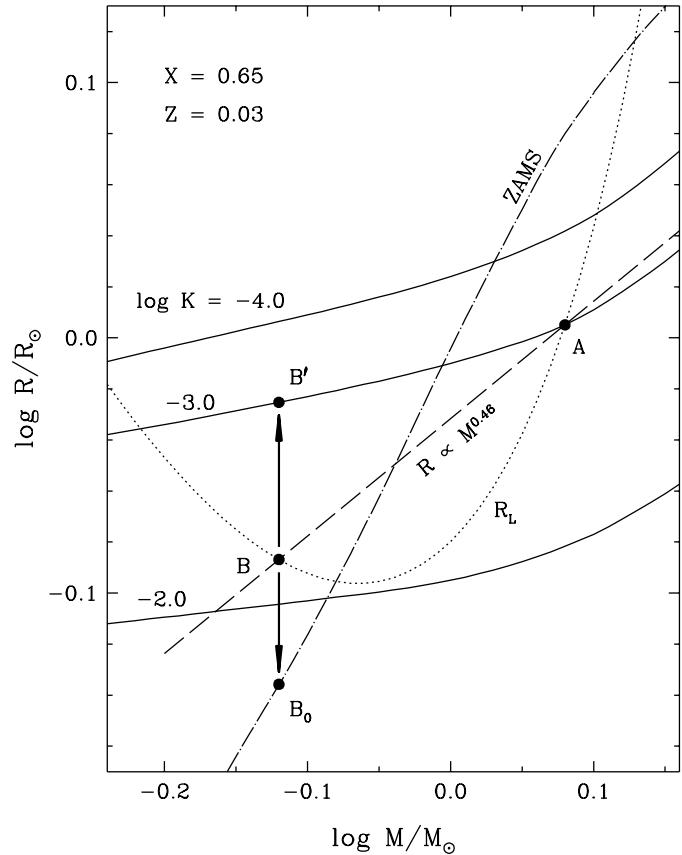}
\caption{The mass-radius diagram for components of W~UMa binaries, after Lucy
(1976).  Constant-$K$ solutions are indicated as in Figure~3.  The zero-age
main sequence (ZAMS) is adopted from Copeland, Jensen, \& J{\o}rgensen (1970). 
The secondary component fills its Roche lobe at point $B$, in marginal contact
with the primary at point $A$.  In this state, it is in hydrostatic
equilibrium, but can exist in thermal equilibrium only at $B^{\prime}$ (in
contact) or at $B_0$ (detached).  In marginal contact, the binary components
evolve along the curve labeled $R_L$, the Roche lobe mass-radius relation for
constant orbital angular momentum.}
\end{figure}
In the contact phase, TRO models successfully reproduce late-type W~UMa-like
systems even with unevolved components (where thermal equilibrium models
fail).  However, in the semi-detached phase (which spans a duration and color
range comparable with the contact phase, with the same range of orbital
periods), the primary and secondary develop very different effective
temperatures.  The TRO model therefore predicts a population of eclipsing
binaries with decidedly unequal minima ($\beta$ Lyrae-, or EB-type binaries)
overlapping in period and color, and similar in number, with the W~UMa
systems.  Examples exist of such short-period EB systems (Lucy \& Wilson
1979), but they appear to be relatively rare, a conclusion emphatically
supported by the fact Rucinski (1998) found only $\sim\!2\%$ (by space
density) of erstwhile contact binaries with $P < 1^{\rm d}$ in the OGLE survey
of Baade's Window displayed poor thermal contact.  Other investigations of TRO
models (Robertson \& Eggleton 1977; Rahunen 1981, 1982, 1983) find that the
amplitude of relaxation oscillations can be damped by systemic angular
momentum losses, but the discrepancy in component temperatures remains a
problem during the semi-detached phase of the oscillations.  Oscillations
might in principle be suppressed completely for sufficiently rapid systemic
angular momentum loss, but the lifetimes of W~UMa binaries then become so
short as to make it impossible to account for their relatively high space density.

\section{Circulation Models}

The difficulties which structural models of W~UMa binaries encounter clearly
require a physical model of mass and energy exchange --- one which can be
implemented in an evolutionary code --- if they are to be resolved.  Over the
years, there have been a number of attempts to build such a model from
hydrodynamical considerations (Hazlehurst \& Meyer-Hofmeister 1973; Moses
1976; Nariai 1976; Webbink 1977; Robertson 1980; Smith \& Smith 1981;
Hazlehurst 1985; K\"{a}hler 1989; Tassoul 1992; Martin \& Davey 1995;
Hazlehurst 1999).  Motivated by the success which Lucy's original hypothesis
(a homogeneous, hydrostatic envelope filling a common equipotential surface)
enjoyed in reproducing the light curves of W~UMa variables, these interaction
models have, with a few exceptions, appealed to a large-scale circulation
between components, driven by small departures from a completely static,
barotropic common envelope.  Since mean vertical temperature gradients ---
those perpendicular to equipotential surfaces --- vastly exceed mean
horizontal temperature gradients, it is clear that advection, not radiative
transport, must be the agent of energy exchange.  In strict hydrostatic
equilibrium, no lateral energy transfer occurs at all.

While it is commonly assumed that the advective flows are only mildly
dynamical, this is by no means assured.  Most light curve solutions for W~UMa
systems yield a very shallow degree of contact, typically of order $\Delta R/R
\approx 0.02$, with the A-type systems tending toward slightly deeper
geometric contact than the W-type systems (\emph{e.g.}, Mochnacki 1981). 
Webbink (1977) found, from a comparison with envelope models for mail sequence
stars in the same mass range, that the observed depths of contact only
slightly exceed the minimum needed to carry the exchange luminosity at
sonic velocity in a closed circulation through the ``neck'' in the
Roche potentials connecting the two stars at the inner Lagrangian ($L_1$)
point (see his Fig. 6).  This estimate is based on the assumption that
radiative and/or convective relaxation across the flow near $L_1$ is
efficient, and that energy exchange is effected by the vertical ascent or
descent of the closed circulation through the vertical entropy gradient in the
common envelope.  Comparison of (i) binary orbital period, (ii) sound travel
time between components at the base of the common envelope, and (iii) local
convective turnover time scale or radiative relaxation time scale (as
appropriate) shows that all three of these time scales are comparable in both
W-type and A-type W~UMa binaries (see Figs. 2 and 3 in Webbink 1977). 
Coriolis effects and energy transport perpendicular to the flow further
complicate efforts to build a conceptual model of energy exchange.

\section{The `Other' Contact Binaries}

The efforts described above to model contact binary stars have concentrated
almost exclusively on the observationally abundant classical W~UMa-type
binaries.  However, binary evolutionary calculations (Benson 1970; Yungel'son
1973; Webbink 1976$b$; Flannery \& Ulrich 1977; Kippenhahn \& Meyer-Hofmeister
1977 --- see especially Nelson \& Eggleton 2001) show unambiguously that the
contact state arises in a far wider range of circumstances than W~UMa systems
reflect, and under conditions (namely, contact between two stars with
radiative envelopes) which cannot be modeled by the usual expedient of
removing luminosity from the primary's surface convection zone and injecting
it into the secondary's.  Massive contact binaries are by no means unknown
(see, \emph{e.g.}, Hilditch \& Bell 1987), but they present a significantly
greater observational challenge than late-type systems in the quest for
reliable spectroscopic orbits and light curve solutions, and so have not
received the same level of attention as W~UMa systems.  Nevertheless, Eggleton
(1996) has suggested that they may in fact be even more common than W~UMa
binaries, in terms of frequency relative to all stars of the same spectral
class.

Structurally, contact binaries with radiative envelopes must still abide by
the precepts laid out above in {\S}2 if they are to achieve complete
hydrostatic equilibrium.  However, the large vertical entropy gradient in a
radiatively stable envelope, large luminosity to be transferred between
components to equilibrate the common envelope, and relatively small heat
capacity of the common envelope (relative to transferred luminosities) all
augur for more vigorously dynamical interaction between components than
typifies their convective counterparts.  Among the W~UMa binaries, the
difficulty or impossibility of achieving thermal equilibrium limits them to
small depths of contact, as argued above; over nuclear time scales, the
evolutionary expansion of the more massive component can then only be
accommodated by expansion of the binary.  That expansion requires mass
transfer from less-massive to more-massive component.  The limited angular
momentum available is gradually subsumed in rotation of the primary star, and
the binary coalesces (Webbink 1976$a$; Rasio 1995).  In massive contact
binaries, the much greater inefficiency of energy exchange between components
opens the possibility that they might, for example, succeed in passing through
deep contact and reverse their initial mass ratios, as do classical Algol
binaries, or otherwise suffer some very different fate from their low-mass
cousins.  The lack of physically plausible models for energy and mass exchange
in radiative contact binaries is a major obstacle in understanding massive
binary evolution.

\section{Conclusions}

The essential difficulty in modeling energy exchange in contact binaries is
that some departure from strict hydrostatic equilibrium is necessary to effect
energy transfer.  But the nature of said departure, and the advective flows
resulting from it, depend sensitively on the thermal state of the respective
stellar envelopes.  In turn, the thermal state of those envelopes is
profoundly affected by energy transfer.  These are not separable issues, and
one needs to be able to model mean dynamical flows, interacting with the
ambient radiation field, as they evolve on the much longer global thermal time
scales of the stellar components.

\acknowledgements

Friends and colleagues too numerous to name have helped shape the opinions
expressed here, but I wish to thank in particular Peter Eggleton and the late
John Whelan for many insightful discussions of contact binary structure, and
Paul Hendry and Stefan Mochnacki for permission to reprint one of their
stunning doppler images of VW Cep (Figure~2).  This work was supported in part
by NASA grant NAG 5-11016.

\end{document}